\UseRawInputEncoding 
\documentclass[twocolumn,superscriptaddress,amssymb,amsmath,floatfix,citeautoscript]{revtex4-1}

\usepackage[pdftex]{graphicx}
\usepackage{amsmath}
\usepackage{color}
\usepackage{natbib}
\usepackage{mathtools}
\usepackage{physics}

\begin{document}

\title{Optical vortex manipulation for topological quantum computation}

\author{Chengyun Hua}
\email{
To whom correspondence should be addressed. E-mail: huac@ornl.gov}
\affiliation{Materials Science and Technology Division, Oak Ridge National Laboratory, Oak Ridge, TN 37831, USA}
\author{G\'abor B. Hal\'asz}
\affiliation{Materials Science and Technology Division, Oak Ridge National Laboratory, Oak Ridge, TN 37831, USA}
\author{Eugene Dumitrescu}
\affiliation{Computational Sciences and Engineering Division, Oak Ridge National Laboratory, Oak Ridge, TN 37831, USA}
\author{Matthew Brahlek}
\affiliation{Materials Science and Technology Division, Oak Ridge National Laboratory, Oak Ridge, TN 37831, USA}
\author{Benjamin Lawrie}
\affiliation{Materials Science and Technology Division, Oak Ridge National Laboratory, Oak Ridge, TN 37831, USA}

\date{\today}

\begin{abstract}

Topological quantum computation based on Majorana bound states may enable new paths to fault-tolerant quantum computing. Several recent experiments have suggested that the vortex cores of topological superconductors, such as iron-based superconductors, may host Majorana bound states at zero energy. However, quantum computation with these zero-energy vortex bound states requires a precise and fast manipulation of individual vortices which is difficult to do in a scalable manner. To address this issue, we propose a control scheme based on local heating via, for example, scanning optical microscopy to braid vortex-bound Majorana zero modes in a two-dimensional topological superconductor. First, we derive the conditions required for transporting a single vortex between two defects in the superconducting material by trapping it with a hot spot generated by local optical heating. Equipped with critical conditions for the vortex motion, we then establish the ideal material properties for vortex braiding and describe how transition errors resulting from finite speed and/or temperature can be minimized. Our work paves the way toward optical or microscopic control of zero-energy vortex bound states in two-dimensional topological superconductors. \footnote{This work was supported by the U. S. Department of Energy, Office of Science, Basic Energy Sciences, Materials Sciences and Engineering Division. This manuscript has been authored by employees of UT-Battelle, LLC under Contract No.~DE-AC05-00OR22725 with the U.S. Department of Energy. The U.S. Government retains and the publisher, by accepting the article for publication, acknowledges that the U.S. Government retains a nonexclusive, paid-up, irrevocable, worldwide license to publish or reproduce the published form of this manuscript, or allow others to do so, for U.S. Government purposes.}

\end{abstract}

\maketitle

Topological quantum systems utilizing superpositions of quantum states encoded in so-called qubits may enable fault-tolerant quantum computing\cite{nayak_non-abelian_2008, sarma_majorana_2015}. State-of-the-art qubit platforms include superconducting Josephson junctions \cite{Krantz2019}, trapped ions \cite{Cirac1995}, and photonic networks \cite{Kok2007}. These qubit technologies have matured sufficiently that small scale quantum computations, albeit with significant noise present, have already been performed. However, a sizable physical qubit overhead is needed to scale architectures and suppress logical error rates for a fault-tolerant architecture. 

An alternative route to encoding high-fidelity logical qubits is to utilize topological modes realized naturally in condensed matter platforms. For example, emergent Majorana zero modes (MZMs) possessing non-Abelian exchange statistics are naturally immune to local decoherence effects due to their fundamentally non-local nature \cite{nayak_non-abelian_2008}. The vortex cores of topological superconductors, such as iron-based superconductors \cite{xu_topological_2016,wu_topological_2016, wang_topological_2015}, may host such MZMs appearing as zero-energy vortex bound states \cite{machida_zero-energy_2019,fu_superconducting_2008,ivanov_non-abelian_2001}. Topological quantum computation then relies on manipulating the positions of the zero-energy vortex bound states to braid or fuse \cite{Bonderson2008} the MZMs. 

The generation and manipulation of individual superconducting vortices has been demonstrated using a range of experimental techniques. Magnetic fields, thermal gradients, and electrical currents have been used to tune the averaged properties of vortex matter\cite{curran_tuning_2018,shi_vortex_2020}. Optical quenching-assisted fast switching of vortex cores has been demonstrated and visualized in situ by Lorentz electron microscopy \cite{fu_optical_2018}. Vortex motion has been manipulated by Lorentz forces using electrical currents \cite{de_souza_silva_controlled_2006,kalisky_dynamics_2009,embon_probing_2015}, by altering the pinning landscape \cite{villegas_superconducting_2003,gutierrez_strong_2007,llordes_nanoscale_2012}, by magnetic forces in magnetic force microscopy \cite{dremov_local_2019, november_scheme_2019, auslaender_mechanics_2009,straver_controlled_2008,polshyn2019manipulating} and scanning superconducting quantum interference device microscopy \cite{kremen_mechanical_2016,kalisky_behavior_2011,gardner_manipulation_2002}, and by local heating using a scanning tunneling microscope \cite{ge_nanoscale_2016} or a far-field optical method \cite{veshchunov_optical_2016}. Despite all these efforts, a theoretical framework for analyzing the dynamics of an individually controlled superconducting vortex is still critically needed. 

\begin{figure*}
\centering
\includegraphics[scale = 0.25]{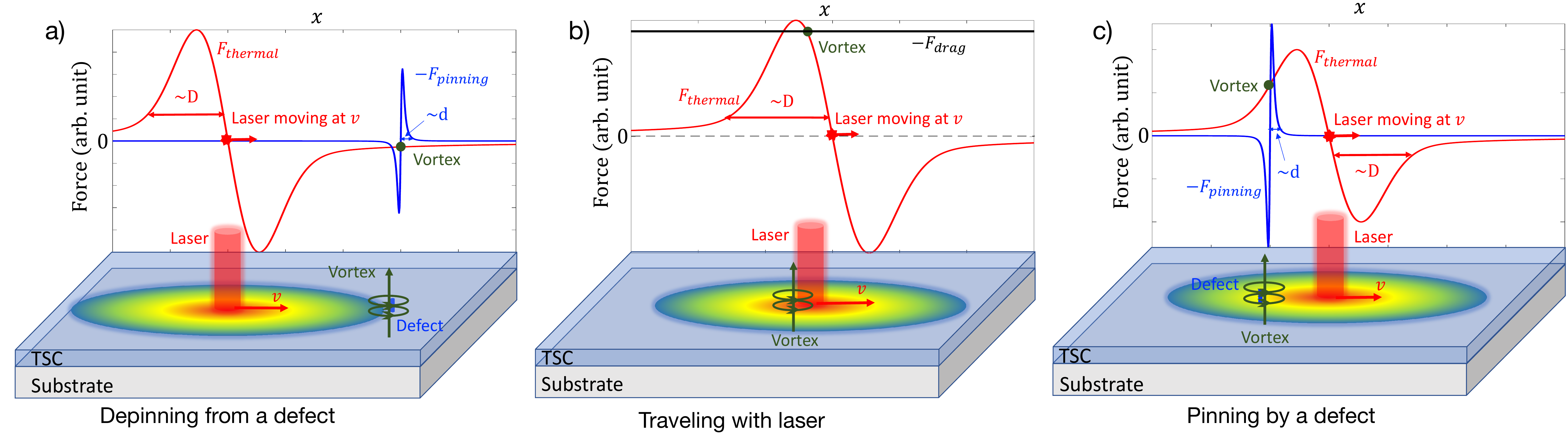}
\caption{General scheme for using a localized optical heating to transport a single superconducting vortex between two stationary defects: (a) picking up the vortex from a defect; (b) transporting the vortex across the superconductor; and (c) delivering the vortex to another defect. At each step, the vortex motion is governed by at most three forces: a time-dependent thermal force resulting from optical heating ($F_{\mathrm{thermal}}$), a stationary pinning force due to the defect ($F_{\mathrm{pinning}}$), and a viscous drag force ($F_{\mathrm{drag}}$).} 
\label{Figure1}
\end{figure*}

In this Letter, we propose a scheme for the use of local heating, based on a scanning laser excitation, to manipulate individual superconducting vortices by picking them up from defects in the material, transporting them across the superconductor, and delivering them to other defects (see Fig.~\ref{Figure1}). We establish the critical conditions required for each step of the process by solving the vortex equation of motion in the presence of a time-dependent thermal field, a stationary pinning potential, and a viscous drag force. Using these universal critical conditions, we provide guidelines for the optimal properties of potential candidate materials and describe how spurious transitions resulting from finite vortex speeds can be minimized.

\emph{General setup.---}We consider a thin slab of type-II superconducting material in a perpendicular magnetic field $H$. In this geometry, magnetic vortices are present below the lower critical field $H_{c1}$. However, for $H \ll H_{c1}$, the mean separation between the vortices, $R \sim \sqrt{\Phi_0 / H}$ (where $\Phi_0 = h/(2e)$ is the flux quantum), is much larger than the penetration depth $\lambda$. We assume that, in the absence of optical heating, these well-separated vortices are localized at appropriate ``pinning sites'' at which superconductivity is suppressed within a region of size $r$. The pinning sites may be intrinsic defects of the material, such as vacancies or impurities, corresponding to $r \sim a$ (where $a$ is the lattice constant), or ``artificial defects'', such as cylindrical voids or electrically gated regions, corresponding to $r \gg a$. We hypothesize that a single vortex can be moved between different pinning sites by dragging it around with a ``hot spot'' resulting from localized optical heating. With such elementary control of magnetic vortices hosting MZMs, one can then implement the Clifford gate set on the topological qubits encoded in these MZMs.

To understand the feasibility of the vortex manipulation process, we aim to establish the general conditions required for the three important steps depicted in Fig.~\ref{Figure1}: (a) picking up the vortex from its initial pinning site; (b) transporting the vortex across the bulk of the superconductor; and (c) delivering the vortex to its final pinning site. Assuming the hot spot moves with a constant speed $v$ and meets each pinning site ``head on'', the position of the vortex, $x(t)$, satisfies a one-dimensional equation of motion during each step:
\begin{equation}
0 = - \eta \dot{x} - \frac{dU(x)}{dx} + \gamma \frac{dT(x-vt)}{dx}. \label{Eq:EofM_full}
\end{equation}
The three terms on the right-hand side describe a viscous drag force, a stationary pinning force, and a time-dependent thermal force resulting from the hot spot, respectively, while the left-hand side is set to zero because the inertial mass of the vortex \cite{Suhl1965} is negligibly small at the experimentally relevant length and speed scales. We note that all forces are defined per unit length along the vortex line piercing through the superconducting slab. Using the Bardeen-Stephen model \cite{Bardeen1965, tinkham_introduction_2004}, the viscous drag coefficient is $\eta = \Phi_0^2 / (2\pi \xi^2 \rho_n)$, where $\xi$ is the coherence length, and $\rho_n$ is the normal-state resistivity. The pinning potential $U(x)$ has a characteristic length scale $d$ and a single minimum at $x = 0$ corresponding to the pinning energy per unit length along the vortex line: $U_0 \equiv -U(0)$. For a pinning site of radius $r$, the length scale is expected to be $d \sim \max (\xi, r)$, while the pinning energy (per unit length) for a cylindrical void is shown in the Supplemental Material (SM) to be $U_0 \sim \Phi_0^2 / (\mu_0 \lambda^2 \xi^2) \min (r^2, \xi^2)$, where $\mu_0$ is the vacuum permeability. This result is also consistent with Ref.~\cite{maurer_vortex_1996} which considers larger magnetic fields ($H > H_{c1}$). To estimate the thermal force coefficient $\gamma$, we recognize that the energy of a vortex decreases as a function of the temperature $T$ due to the weakening of superconductivity \cite{veshchunov_optical_2016}. For a strongly type-II superconductor ($\lambda \gg \xi$), the vortex energy is proportional to the density of Cooper pairs and grows linearly with $T_c-T$ \cite{gennes_superconductivity_1999, fossheim_superconductivity_2005}, where $T_c$ is the superconducting critical temperature. The thermal force coefficient is then $\gamma =  \Phi^2_0/(4\pi \mu_0 \lambda_0^2 T_c)\text{ln} \left( \lambda_0/\xi_0\right)$, where $\lambda_0$ and $\xi_0$ are the values of $\lambda$ and $\xi$ at zero temperature, respectively. Finally, the temperature profile of the hot spot, $T(x-vt)$, is calculated from a two-layer heat diffusion model (see the SM) where we assume that (i) a thin superconducting layer is grown on top of an infinitely thick substrate and (ii) the superconducting layer is subject to an optical heating source of diameter $D_0$ moving with a constant speed $v$. This calculation gives a hot-spot temperature profile with a single maximum $\Delta T$ (with respect to the bulk temperature) and a characteristic length scale $D \sim D_0$.

To make the subsequent analysis simpler, we introduce a dimensionless vortex position, $\tilde{x} = x/d$, a dimensionless time, $\tilde{t} = vt/d$ (where $v$ is the hot-spot speed), and dimensionless numbers comparing both the pinning force and the thermal force to the viscous drag force. In terms of these dimensionless variables, the equation of motion can be written as
\begin{equation}
\dot{\tilde{x}}= \alpha f_p(\tilde{x}) + \beta f_{th}\left(\frac{\tilde{x}-\tilde{t}}{\tilde{D}}\right), \label{eq:EofM_dimensionless}
\end{equation}
where $\tilde{D} = D/d$ is the dimensionless hot-spot size, while $\alpha \sim U_0 / (d \eta v)$ and $\beta \sim \gamma \Delta T /(D \eta v)$ are dimensionless ratios of the maximal pinning and thermal forces to the viscous drag force at the hot-spot speed $v$, respectively. The dimensionless functions $f_{p,th}(z)$ are antisymmetric, $f_{p,th}(-z) =-f_{p,th}(z)$, have a single zero, $f_{p,th}(0) = 0$, a single maximum, $f_{p,th}(-1) = 1$, and a single minimum, $f_{p,th}(1)=-1$. We take a Lorentzian shape for the pinning potential, corresponding to $f_p(z)=-16z/(3+z^2)^2$, while the temperature profile of the hot spot is found to resemble a Gaussian shape (see the SM).  We emphasize, however, that our main results do not depend on the precise forms of these functions.

\emph{Critical conditions.---}We now establish the general conditions required for the three steps of the vortex manipulation process in terms of the dimensionless parameters $\alpha$, $\beta$, and $\tilde{D}$. This condition is the simplest for the second step when the vortex is traveling with the hot spot [see Fig.~\ref{Figure1}(b)]. In this case, there is no pinning force, and the first term on the right-hand side of Eq.~(\ref{eq:EofM_dimensionless}) vanishes. Then, since the steady-state solution of Eq.~(\ref{eq:EofM_dimensionless}) must be of the form $\tilde{x}=\tilde{t}+\tilde{x}_0$, where $\tilde{x}_0$ is a constant displacement of the vortex with respect to the center of the hot spot, we can use $\dot{\tilde{x}}=1$ and $f_{th}(\tilde{x}_0/\tilde{D})<1$ to establish that the vortex can only travel with the hot spot if $\beta > 1$. This result has a simple physical interpretation: the maximal thermal force must exceed the viscous drag force at the hot-spot speed.

For the first and third steps [see Figs.~\ref{Figure1}(a) and \ref{Figure1}(c)], the conditions are more complex because all three terms of Eq.~(\ref{eq:EofM_dimensionless}) are present. In each case, the main question is whether the vortex is eventually trapped by the pinning site or carried away by the hot spot. We first show that the vortex always ends up at the pinning site for $\delta \equiv \beta - \alpha < 0$ and at the hot spot for $\delta > 1$. Indeed, regardless of its initial condition (i.e., if it is originally at the pinning site or the hot spot), the vortex can only be carried away by the hot spot if, at some point, it goes through position $\tilde{x} = 1$ with a positive speed $\dot{\tilde{x}} > 0$. According to Eq.~(\ref{eq:EofM_dimensionless}), the speed of the vortex at this critical position is $\dot{\tilde{x}} = -\alpha + \beta f_{th} [(1-\tilde{t})/\tilde{D}] < -\alpha + \beta$. Therefore, the vortex is necessarily trapped by the pinning site if $\alpha > \beta$. Similarly, regardless of its initial condition, the vortex can only end up trapped by the pinning site if, at some point, its relative position with respect to the hot spot, $\tilde{x} - \tilde{t}$, becomes smaller than $-\tilde{D}$ which, in turn, requires a negative relative speed, $\dot{\tilde{x}} - 1 < 0$ at this critical relative position. From Eq.~(\ref{eq:EofM_dimensionless}), the critical relative speed is $\dot{\tilde{x}} - 1 = \alpha f_{p} (\tilde{t}-\tilde{D}) + \beta - 1 > -\alpha + \beta - 1$. Therefore, the vortex is necessarily carried away by the hot spot if $\beta > \alpha + 1$. We emphasize that the statements derived in this paragraph are completely exact.

\begin{figure}
\centering
\includegraphics[scale = 0.4]{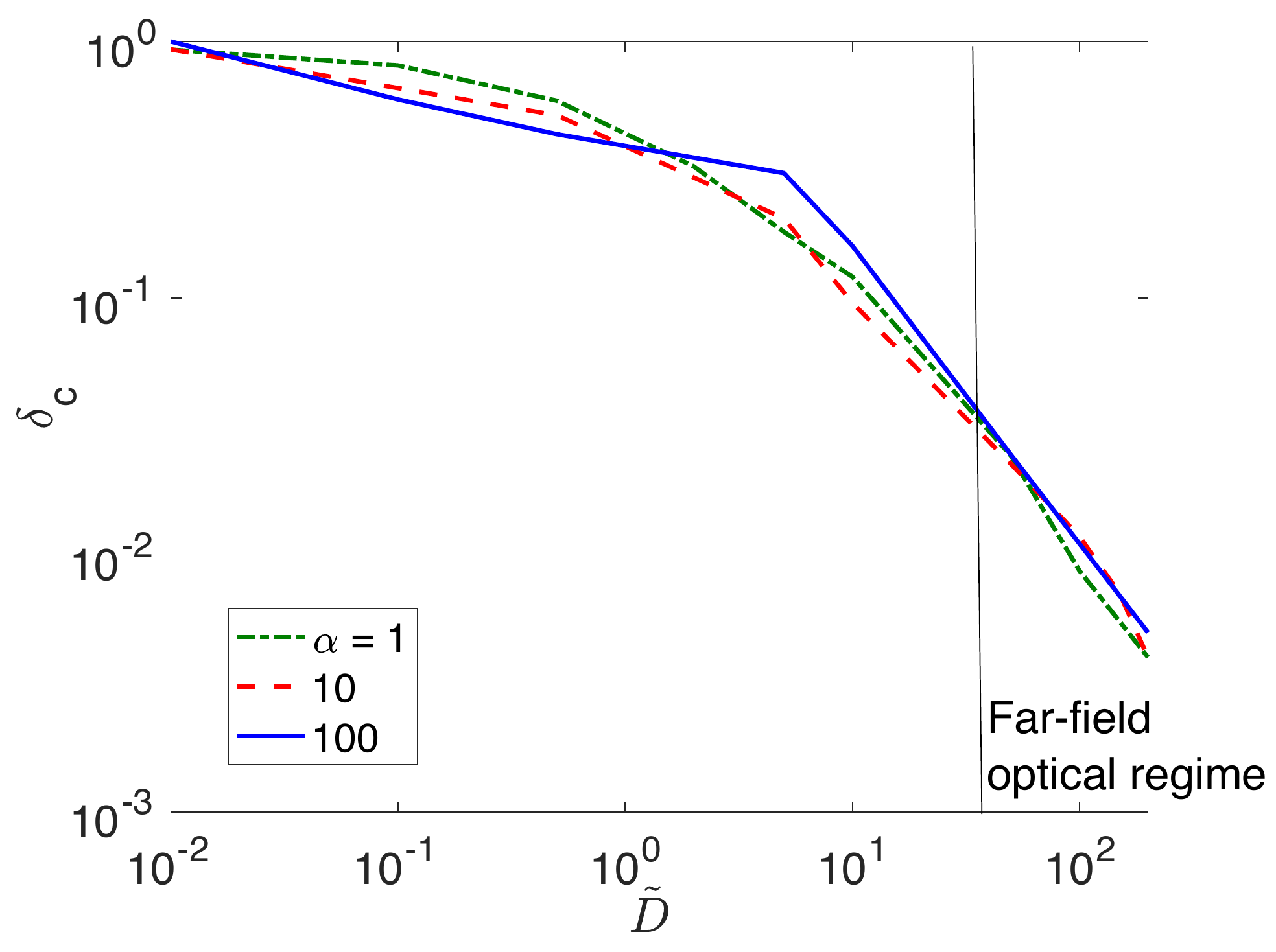}
\caption{Critical value $\delta_c$ of the dimensionless excess thermal force, $\delta = \beta-\alpha$, at which the vortex is carried away by the hot spot rather than being trapped by the pinning site; $\delta_c$ is plotted against $\tilde{D} = D/d$ for dimensionless pinning forces $\alpha = 100$ (blue solid line), $\alpha = 10$ (red dashed line), and $\alpha = 1$ (green dash-dotted line).} \label{Figure2}
\end{figure}

For the intermediate regime, $0 < \delta < 1$, the simple analytical arguments above do not determine the fate of the vortex. In this case, we solve Eq.~(\ref{eq:EofM_dimensionless}) numerically using an adaptive stepsize Runge-Kutta integration scheme \cite{press_adaptive_1992}. Our first main result is that there is a well-defined critical value $\delta_c$ such that, \emph{regardless of its initial condition}, the vortex ends up at the pinning site for $\delta < \delta_c$ and at the hot spot for $\delta > \delta_c$. This critical value $\delta_c$ is plotted as a function of $\alpha$ and $\tilde{D}$ in Fig.~\ref{Figure2}. While its behavior is generally complex for $\tilde{D} \sim 1$, we find two universal limits, corresponding to $\tilde{D} \gg 1$ and $\tilde{D} \ll 1$, in which the critical values are $\delta_c \ll 1$ and $\delta_c \approx 1$, respectively.

These two universal limits have simple physical interpretations. In the $\tilde{D} \gg 1$ limit, the drag force is negligible, and the fate of the vortex is determined by the balance of the maximal pinning and thermal forces. Therefore, the critical condition simply reduces to $\beta = \alpha$. In the $\tilde{D} \ll 1$ limit, the drag force takes a constant value, corresponding to the hot-spot speed, and maximally supports the pinning force against the thermal force. Therefore, the critical condition becomes $\beta = \alpha + 1$. We derive these limiting critical conditions, along with their lowest-order corrections for finite $\tilde{D}$, in the SM.

\begin{figure}
\centering
\includegraphics[scale = 0.2]{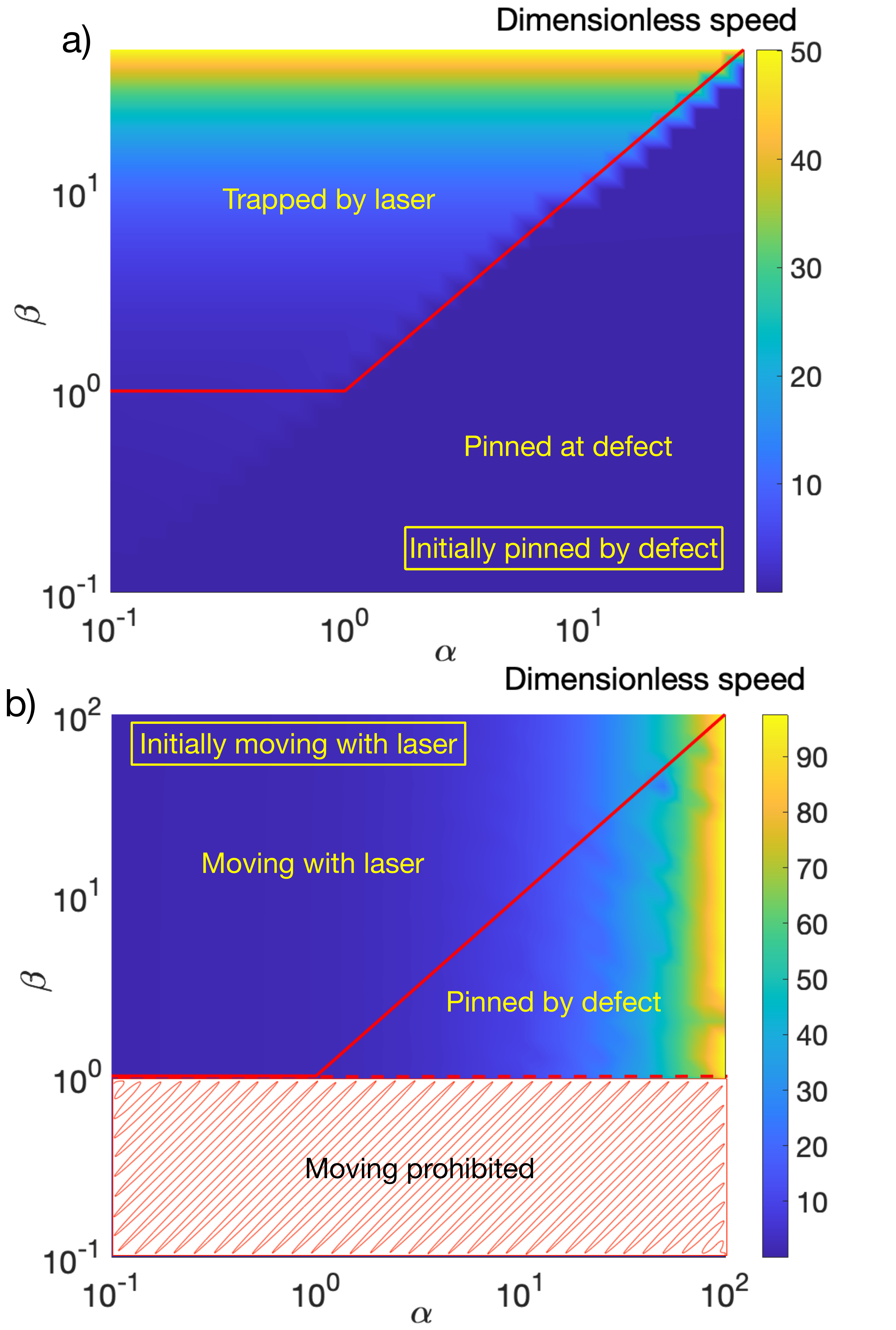}
\caption{Final state of the vortex after the hot spot moves over the pinning site if the vortex is initially (a) at the pinning site or (b) at the hot spot. Red lines separate different final states and correspond to critical conditions in terms of the dimensionless pinning ($\alpha$) and thermal ($\beta$) forces. The shading gives the maximal speed of the vortex during the entire process, while the hatching indicates that the initial state of the vortex is unphysical.} \label{Figure3}
\end{figure}

For a diffraction-limited scanning laser excitation, the first universal limit with $\tilde{D} \gg 1$ is the experimentally relevant one. Our results for the critical conditions in this limit are summarized in Fig.~\ref{Figure3}. If the vortex is originally at the pinning site [see Fig.~\ref{Figure3}(a)], it can be picked up by the hot spot if (i) $\beta > \alpha$ so that the maximal thermal force exceeds the maximal pinning force \emph{and} (ii) $\beta > 1$ so that the thermal force can keep the vortex trapped while moving it. If the vortex is originally at the hot spot [see Fig.~\ref{Figure3}(b)], its initial condition is only physical for $\beta > 1$ in the first place. The vortex can then be delivered to the pinning site if $\alpha > \beta$ so that the maximal pinning force exceeds the maximal thermal force. The simplicity and universality of these critical conditions facilitates the design of concrete experimental schemes for optical vortex manipulation in topological superconductors.

\emph{Material properties.---}To establish the ideal material properties for our scheme, we recognize that the pickup condition $\beta > \alpha$ and the delivery condition $\beta < \alpha$ are mutually exclusive. Therefore, we must be able to tune at least one of the parameters $\alpha$ and $\beta$. For example, $\beta$ can be tuned by adjusting the laser power, while $\alpha$ can be tuned in the case of electrically gated pinning sites. Nevertheless, to minimize the amount of tuning and any disruptions it may cause, it is important to keep the two parameters close to each other: $\beta \approx \alpha$. Using the definitions of $\alpha$ and $\beta$, and neglecting all $O(1)$ and logarithmic factors, this condition immediately translates to
\begin{equation}
\frac{\beta} {\alpha} \sim \frac{\gamma \Delta T} {U_0} \frac{d} {D} \sim \frac{\max(\xi^3, r^3)} {r^2 D} \frac{\Delta T} {T_c} \sim 1. \label{eq-mat-1}
\end{equation}
Because of $\Delta T < T_c$ and the large size of the hot spot, $D \sim 1$ $\mu$m, we need sufficiently small pinning sites, $r \ll \xi$, to satisfy Eq.~(\ref{eq-mat-1}). In this regime, the ideal pinning-site radius is then $r \sim \sqrt{\xi^3 \Delta T / (D T_c)}$. Moreover, since $r \gtrsim a$ for any reasonable pinning site, we need a sufficiently large coherence length:
\begin{equation}
\xi \gtrsim \xi_c \sim \left( a^2 D \frac{T_c} {\Delta T} \right)^{1/3}. \label{eq-mat-2}
\end{equation}
In particular, we need $\xi \sim \xi_c$ to use the intrinsic defects of the material, such as vacancies or impurities, and $\xi \gg \xi_c$ to use artificial defects that are much stronger than these intrinsic defects. For $a \sim 1$ nm and $D \sim 1$ $\mu$m, the critical coherence length is $\xi_c \sim 10$ nm (assuming $\Delta T \sim T_c$).

\emph{Diabatic errors.---}In addition to satisfying the critical conditions, it is important to make the vortex manipulation process as smooth as possible to maximize the topological protection for the MZMs hosted by the vortices. Generally, the ``diabatic errors'' \cite{Cheng2011, Knapp2016} resulting from non-adiabatic (finite-speed) operation are expected to be small as long as the typical time scale $\tau$ characterizing the operation is much larger than the inverse of the superconducting gap $\Delta$ \footnote{While localized states may exist inside the bulk gap, diabatic transitions into these localized states are not detrimental to the quantum information (see, e.g., Ref.~\cite{Cheng2011}).}. Using $\tau \sim \xi / v$ and $\xi \sim v_F / \Delta$, where $v_F$ is the Fermi speed, this condition translates to $v \ll v_F$. For realistic hot-spot speeds, $v \sim 1$ cm/s, this condition is then easily satisfied for any reasonable material while the vortex is traveling in the bulk of the superconductor. However, as shown by Fig.~\ref{Figure3}, the speed of the vortex can be orders of magnitude larger when it is picked up by the hot spot or delivered to a pinning site. To minimize these maximal speeds during the vortex manipulation process, it is important to keep $\beta$ as close to $1$ as possible (see Fig.~\ref{Figure3}), which requires
\begin{equation}
\beta \sim \frac{\gamma \Delta T} {D \eta v} \sim \frac{\xi^2} {\lambda^2} \frac{\Delta T} {T_c} \frac{\rho_n} {\mu_0 D v} \gtrsim 1. \label{eq-mat-3}
\end{equation}
For $D \sim 1$ $\mu$m, $v \sim 1$ cm/s, and $\rho_n \sim 10^{-7}$ $\Omega$m, we obtain $\rho_n/(\mu_0 D v) \sim 10^7$. Therefore, we need to minimize $\beta$ as much as possible by using an extreme type-II superconductor ($\lambda / \xi \gg 1$) with a small normal-state resistivity $\rho_n$. Importantly, however, even for small speeds, $v \ll v_F$, the diabatic errors resulting from finite speed are not exponentially suppressed by any kind of topological protection \cite{Knapp2016}. Their precise effect on the MZMs during our vortex manipulation process will be investigated in a future work.

\emph{Conclusion.---}We have presented a scheme for the use of local heating by a scanning laser excitation to manipulate Majorana bound states emergent in the vortex cores of topological superconductors. In the practically relevant regime, the conditions required for transporting a single vortex between two defects in the material are universal and surprisingly simple. In particular, the vortex can be moved across the bulk of the superconductor if the maximal thermal force exceeds the viscous drag force, while it can be picked up from (delivered to) a defect if the maximal thermal force is larger (smaller) than the maximal pinning force. We have established the ideal material properties for the implementation of our scheme by considering both intrinsic and artificial defects.

Although our formalism does not explicitly account for the Majorana bound states inside the vortex cores, the nonlocal encoding of quantum information guarantees its topological protection. Specifically, the decoherence rate is exponentially suppressed as long as the separation between any two vortices is much larger than the coherence length $\xi$ and the maximum temperature due to local laser heating is much smaller than the bulk gap $\Delta$. While localized states may exist inside the gap, it is known that transitions into such localized states are not detrimental to the quantum information encoded in the Majorana bound states \cite{Akhmerov2010}. We have also argued that the vortex motion is close to adiabatic for realistic speeds of the laser excitation and that any diabatic errors resulting from finite speed are minimized for extreme type-II superconductors. The results of our work provide much needed theoretical guidance on optimizing control over superconducting vortices using local heating and will facilitate the design of concrete experimental protocols for precise and rapid vortex manipulation.



\end{document}